\documentclass[showpacs,amsmath,amssymb,twocolumn,superscriptaddress]{revtex4}

\usepackage{graphicx}% Include figure files
\usepackage{dcolumn}% Align table columns on decimal point
\usepackage{bm}% bold math
\usepackage{ulem}

\begin{document}

\setlength{\abovedisplayskip}{3pt plus1pt minus1pt}
\setlength{\belowdisplayskip}{3pt plus1pt minus1pt}
\setlength{\textfloatsep}{10pt plus1pt minus1pt}
\setlength{\abovecaptionskip}{0pt}
%\preprint{APS/123-QED}

\draft

\title{Isospin effect in the statistical sequential decay}
\author{W. D. Tian}
\email{tianwendong@sinap.ac.cn}
\author{Y. G. Ma}
\author{X. Z. Cai}
\author{D. Q. Fang}
\author{W. Guo}
\author{W. Q. Shen}
\author{K. Wang}
\author{H. W. Wang}
\affiliation{Shanghai Institute of Applied
Physics, Chinese Academy of Sciences, P. O. Box
800-204, 201800, Shanghai, China}

\author{M. Veselsky}
\affiliation{Institute of Physics, Slovak Academy of Sciences,
Dubravska cesta 9, Bratislava, Slovakia}
\date{\today}

\begin{abstract}
Isospin effect of the statistical emission fragments from the
equilibrated source is investigated in the frame of statistical
binary decay implemented into GEMINI code, isoscaling behavior is
observed and the dependences of  isoscaling parameters $\alpha$
and $\beta$ on emission fragment size, source size, source isospin
asymmetry and excitation energies are studied. Results show that
$\alpha$ and $\beta$ neither depends on light fragment size nor on
source size. A good linear dependence of $\alpha$ and $\beta$ on
the inverse of temperature $T$ is manifested  and   the
relationship of
$\alpha=4C_{sym}[(Z_{s}/A_{s})_{1}^{2}-(Z_{s}/A_{s})_{2}^{2}]/T$
and
$\beta=4C_{sym}[(N_{s}/A_{s})_{1}^{2}-(N_{s}/A_{s})_{2}^{2}]/T$
from different isospin asymmetry sources are satisfied. The
symmetry energy coefficient $C_{sym}$ extracted from simulation
results is  $\sim$ 23 MeV which includes both the volume and
surface term contributions, of which the surface effect seems to
play a significant role in the symmetry energy.
\end{abstract}


\maketitle

\section{\label{sec:sec1}Introduction}
The growing interest in isospin effects in nuclear reactions is
motivated by an increasing awareness of the importance of the
symmetry term in the nuclear equation of state. The availability
of beams with large neutron-to-proton ratio, $N/Z$, provides the
opportunity to explore the symmetry energy in very
isospin-asymmetric nuclear systems. In such reactions, isospin
degree of freedom has prominent roles and can be served as a
valuable probe of the symmetry energy term of the nuclear equation
of state. The isotopic composition of the nuclear reaction
products contains important information on the role of the isospin
on the reaction process. $N/Z$ degree of freedom and its
equilibration, as well as the isospin asymmetry dependent terms of
the nuclear equation of state (EOS) \cite{To99, Li98, Ma99, Mu95,
Bo94, Gu01}, have motivated detailed measurements of the isotopic
distributions of reaction products.

One important observable in heavy-ion collisions for determining
the symmetry energy experimentally is the fragment isotopic
composition investigated with the recently developed isoscaling
approach \cite{Xu00, Ts01}. The isoscaling approach attempts to
isolate the effects of the nuclear symmetry energy in the fragment
yields, thus allowing a direct study of the symmetry energy term
in the nuclear binding energy during formation of hot fragments.
Isoscaling refers to a general exponential relation between the
yield ratios of  given fragments between two reactions which
differ only in their isospin asymmetry $(N/Z)$. In particular, if
two reactions, 1 and 2, lead to primary fragments having
approximately same temperature but different isospin asymmetry,
the ratio $R_{21}(N,Z)$ of the yields of a given fragment $(N,Z)$
from these primary fragments exhibits an exponential dependence on
the neutron number $N$ and the atomic number $Z$ by following
form:
\begin{equation}
R_{2}(N,Z) = \frac{Y_{2}(N,Z)}{Y_{1}(N,Z)} = C\exp(\alpha N+\beta
Z), \label{eq:eq1}
\end{equation}
where $\alpha$ and $\beta$ are two scaling parameters and $C$ is
an overall normalization constant. This scaling behavior has been
observed in a very broad range of reactions \cite{Br93, Vo78,
Ve04, So03, Bo02, Ge04, Sh04, Fe05, Ig06, Ko07} and theoretical
calculations
\cite{Ba05,Ts01a,Ts01b,Wa05,Ma05,Ma04,On03,Ti05,Ra05,Do06,Zh06,Ti06,Fa06}.
A review paper can be found in recent international "World
Consensus Initiative" book on "Dynamics and Thermodynamics with
Nuclear Degrees of Freedom" \cite{Colonna}.

The aim of the present paper is to investigate the isospin effect
in sequential binary decay implemented into GEMINI code by
isolating formation of the excited composite system, to learn the
sensitivity of the isoscaling parameters with respect to the
observable characterizing the source's state by only observing the
first step decayed fragments, such as $N/Z$ distribution and
isotopic characteristics of the evaporated light particle.
Sequential binary decay model GEMINI \cite{Ch88, Gemini, Ch97} has
been successfully used to describe the light particle evaporation,
complex fragment emission, and $N/Z$ distribution of the
equilibrated compound source.

The article is organized as follows. Section \ref{sec:sec2} makes a
simple review on the sequential binary decay implemented into GEMINI
code, where a brief description of the symmetry energy adopted in
the binding energy calculation and the method to extract temperature
are given. In Section \ref{sec:sec3}, isoscaling phenomenon and its
dependence on compound source size, excitation energy and isospin
asymmetry are presented, the sensitivities of the isoscaling
parameters $\alpha$ and $\beta$ to observables characterizing the
state of the source are discussed in detail. The contribution from
surface effect on the symmetry energy term is discussed in Section
\ref{sec:sec4}. Finally a summary is given.

\section{\label{sec:sec2}Model Overview}

GEMINI model \cite{Ch88,Gemini} calculates the decay of compound
nuclei by modes of sequential binary decays. All possible binary
divisions from light-particle emission to symmetric division are
considered. The model employs a Monte Carlo technique to follow
the decay chains of individual compound nuclei through sequential
binary decays until the resulting products are unable to undergo
further decay.

The decay width for the evaporation of fragments with $Z\le 2$ is
calculated using the Hauser-Feshbach formalism \cite{Ha52}. For
the emission of a light particle $(Z_{1},A_{1})$ with spin $J_{1}$
from a system $(Z_{0},A_{0})$ with excitation energy $E^{*}$ and
spin $J_{0}$, leaving the residual system $Z_{2},A_{2}$ with spin
$J_{2}$, the decay width is given by:
\begin{eqnarray}
&
&\Gamma(Z_{1},A_{1},Z_{2},A_{2})=\frac{2J_{1}+1}{2\pi\rho_{0}}
\sum_{l=|J_{0}-J_{2}|}^{J_{0}+J_{2}} \nonumber\\
& &\int_{0}^{E^{*}-B-E_{rot}(J_{2})}
T_{l}(\varepsilon)\rho_{2}(U_{2},J_{2})d\varepsilon.
\label{eq:eq2}
\end{eqnarray}
In this equation $l$ and $\varepsilon$ are the
orbital angular momentum and kinetic energy of
the emitted particle, $\rho_{2}(U_{2},J_{2})$ is
the level density of the residual system with
thermal excitation energy
\begin{equation}
U_{2}=E^{*}-B-E_{rot}(J_{2})-\varepsilon,
\label{eq:eq3}
\end{equation}
where $B$ is the binding energy, $E_{rot}(J_{2})$
is the rotation plus deformation energy of the
residual system, $\rho_{0}$ is level density of
the initial system and $T_{l}$ is the
transmission coefficients.

For binary divisions corresponding to the emission of heavier
fragment, the decay width is calculated using the transition state
formalism of Moretto \cite{Mo75}
\begin{equation}
\Gamma(Z_{1},A_{1},Z_{2},A_{2})=\frac{1}{2\pi\rho_{0}}
\int_{0}^{E^{*}-E_{sad}(J_{0})}\rho_{sad}(U_{sad},J_{0})d\varepsilon,
\label{eq:eq4}
\end{equation}
where $U_{sad}$ and $\rho_{sad}$ are the thermal energy and level
density of the conditional saddle-point configuration,
respectively,
\begin{equation}
U_{sad}=E^{*}-E_{sad}(J_{0})-\varepsilon,
\label{eq:eq5}
\end{equation}
where $E_{sad}$ is the deformation plus rotation energy of the
saddle-point configuration and $\varepsilon$ now is the kinetic
energy of the translational degree of freedom.

The symmetry energy term due to the neutron-proton excess is
presented in calculating the  masses of nuclei. For heavy systems
(Z$>$12), the masses of the initial and residual systems are
obtained from the Yukawa-plus-exponential model of Krappe, Nix and
Sierk \cite{Kr79} without the shell correction, pairing correction
term for odd-odd nuclei is included. The parameters for this model
are taken from the fit to experimental masses of M\"{o}ller and Nix
\cite{Mo81}. For very light systems (A$\le$12), masses of the nuclei
are calculated from the experimental ones.

Taking into account the effect of a predicted increase in the
symmetry energy associated with the temperature dependence of
effective nucleon mass in the surface of the nucleus \cite{Do94},
the kinetic part of the symmetry energy is related to the level
spacing at the Fermi surface and so it is also related to the
level density parameter $a_{T}$, the following
temperature-dependent kinetic symmetry energy was therefore
included to calculate the temperature dependent level density
parameter $a_{T}$ in the GEMINI simulations \cite{Ch97, Le95}
\begin{equation}
E_{sym}^{kin}(T)=0.82247(\frac{1}{a_{T}}-\frac{1}{a_{0}})(N-Z)^{2},
\label{eq:eq6}
\end{equation}
where $a_{0}=(1.6A+1.8A^{2/3})/15.5$.

The effective thermal excitation energy of the equilibrated system
before light particle evaporation and saddle-point configuration
before heavy fragment emission can be got by
\begin{equation}
E_{ex}^{ther}=U_{2}-E_{sym}^{kin}, \label{eq:eq7}
\end{equation}
or
\begin{equation}
E_{ex}^{ther}=U_{sad}-E_{sym}^{kin}. \label{eq:eq8}
\end{equation}

Then the nuclear system temperature is approximately
\begin{equation}
T=\sqrt{E_{ex}^{ther}/a_{T}}. \label{eq:eq9}
\end{equation}

\section{\label{sec:sec3}Isoscaling Behaviors}

\begin{figure}
\includegraphics[width=0.5\textwidth]{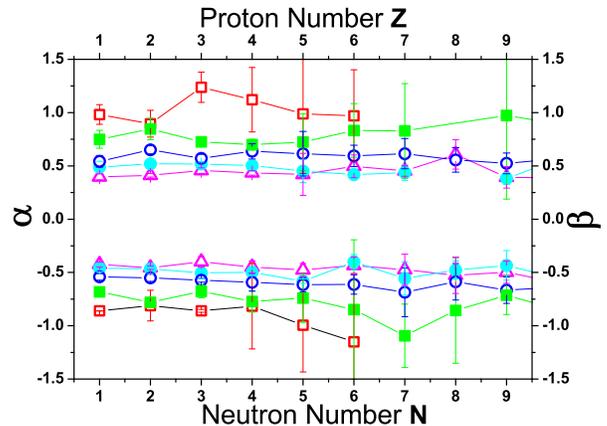}
\caption{\label{fig:fig1} (Color online) Isoscaling parameters
$\alpha$ (positive values) and $\beta$ (negative values) as a
function of the fragment proton number $Z$ or neutron  number $N$
from source pair ($Z_s$ = 50, $A_s$ = 100) and ($Z_s$ = 50, $A_s$ =
105) at various excitation energies: $E_{ex}$ = 1.0 (open squares),
1.4 (solid squares), 2.0 (open circles), 2.4 (solid circles), 3.0
(open up triangles) MeV/nucleon. }
\end{figure}

\begin{figure}
\includegraphics[width=0.5\textwidth]{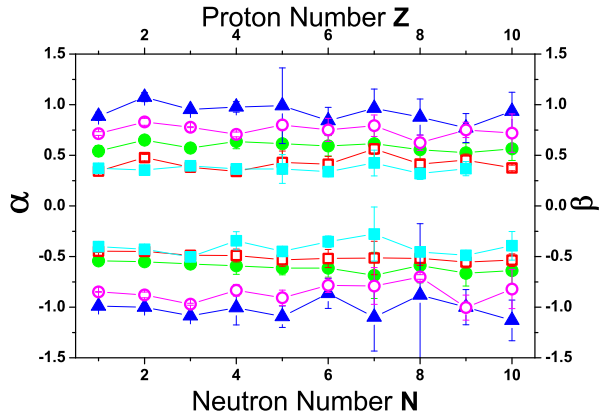}
\caption{\label{fig:fig2} (Color online) Isoscaling parameters
$\alpha$ (positive values) and $\beta$ (negative values) as a
function of the fragment proton number $Z$ or neutron  number $N$
from source pairs with the fixed proton number $Z_s$ =50 at
excitation energies $E_{ex}$ = 2 MeV/nucleon. Symbols in figure
correspond to $Y_{A_s=115}/Y_{A_s=110}$ (solid squares),
$Y_{A_s=110}/Y_{A_s=105}$ (open square), $Y_{A_s=105}/Y_{A_s=100}$
(solid circles), $Y_{A_s=115}/Y_{A_s=105}$ (open circles),
$Y_{A_s=110}/Y_{A_s=100}$ (solid up triangles).}
\end{figure}

\begin{figure}
\includegraphics[width=0.5\textwidth]{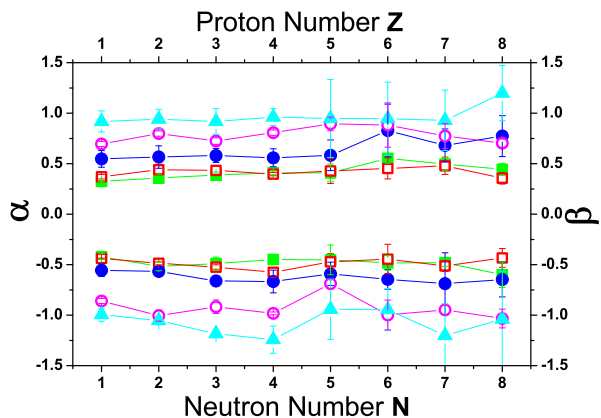}
\caption{\label{fig:fig3} (Color online) Same as Fig. 2 but for
source pairs with the fixed proton number $Z_s$ =30 at excitation
energies $E_{ex}$ = 2 MeV/nucleon. Symbols in figure correspond to
$Y_{A_s=69}/Y_{A_s=66}$ (solid squares), $Y_{A_s=66}/Y_{A_s=63}$
(open square), $Y_{A_s=63}/Y_{A_s=60}$ (solid circles),
$Y_{A_s=69}/Y_{A_s=63}$ (open circles), $Y_{A_s=66}/Y_{A_s=60}$
(solid up triangles).}
\end{figure}

To make a systematic study of source parameters which might
influence the isoscaling behavior, in our present work several
pairs of equilibrated sources are considered at various initial
excitation energies $E_{ex}$ = 1.0, 1.4, 2.0, 2.4 and 3.0
MeV/nucleon. To avoid possible effects of different magnitudes of
Coulomb interaction on isotopic distributions, we consider pairs
of sources with the same proton numbers $Z_s$ but different mass
numbers $A_s$. The equilibrated source pairs are chosen at
different mass region and system isospin asymmetry $N/Z$, two
groups of the  source pairs have been used: (1) $Z_s$ = 50 with
$A_s$ = 100, 105, 110 and 115, respectively; (2) $Z_s$ = 30 with
$A_s$ = 60, 63, 66 and 69, respectively. We adopt the widely used
convention to denote with the index
$^{\prime\prime}$2$^{\prime\prime}$ the more neutron-rich system
and with the index $^{\prime\prime}$1$^{\prime\prime}$ the more
neutron-poor system. In this situation the value of $\alpha$ is
always positive because more neutron-rich clusters will be
produced by the neutron-richer source and the value of $\beta$ is
always negative. The yield ratios $R_{21}(N,Z)$ are calculated and
the corresponding isoscaling behaviors are investigated over all
possible decayed fragments. In the present study, we will focus on
the first decay step  to investigate the decay products even
though a real decay chain is usually longer than one for the
deexcitation process of  a thermal source. The advantage of
studies on the first step is that the source size and temperature
parameters are well defined.

Fig. \ref{fig:fig1} plots the isoscaling parameters $\alpha$ and
$\beta$ as a function of fragment proton number $Z$ and neutron
number $N$ for source pairs with fixed proton number: ($Z_s$ = 50,
$A_s$ = 100 ($N_s/Z_s$ = 1.0) and $A_s$ = 105 ($N_s/Z_s$ = 1.1))
at excitation energies 1.0, 1.4, 2.0, 2.4 and 3.0 MeV/nucleon.
Fig. \ref{fig:fig2} and Fig. \ref{fig:fig3} plot the isoscaling
parameters $\alpha$ and $\beta$ as a function of fragment proton
number $Z$ and neutron number $N$ for different size source pairs
at excitation energies 2.0 MeV/nucelon respectively. Fig.
\ref{fig:fig2} is the source pairs with source atomic number $Z_s$
= 50 and different mass number $A_s$ = 100, 105, 110 and 115, and
Fig. \ref{fig:fig3} is for the source pairs with source atomic
number $Z_s$ = 30 and different mass number $A_s$ = 60, 63, 66 and
69.

Fig. \ref{fig:fig1}, \ref{fig:fig2} and \ref{fig:fig3} show that the
values of isoscaling parameters $\alpha$ and $\beta$ are essentially
flat with the fragment proton number $Z$ or neutron number $N$ in
the case ploted, as well as cases which included in the simulations
which were not plotted in the figure samples. Average isoscaling
parameters $\alpha$ and $\beta$ can be calculated over the flat
region to discuss the dependence of $\alpha$ and $\beta$ on the
properties of emission source, such as excitation energy, source
size and source asymmetry of the isospin. The average $\alpha$ is
calculated over the range $Z \le 9$ and the average $\beta$ is
calculated over the range $N \le 9$, which keep both average values
are calculated over on the same fragments region. In the following
sections isoscaling parameters $\alpha$ and $\beta$ refer to the
calculated average values.

\subsection{Dependence of $\alpha$ and $\beta$ on excitation energy}

\begin{figure}
\includegraphics[width=0.5\textwidth]{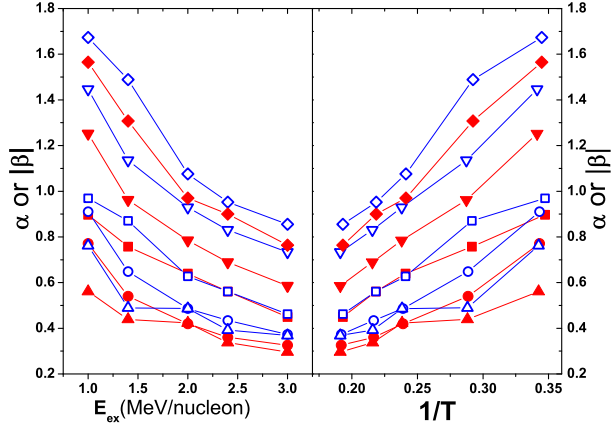}
\caption{\label{fig:fig4} (Color online) Dependence of isoscaling
parameters $\alpha$ and $|\beta|$ on excitation energy (left
panel) and the inverse temperature $1/T$ (right panel) for various
source pairs with same proton number $Z_s$ = 30. Symbols in the
figure are $\alpha$ (solid symbols) or $|\beta|$ (open symbols)
from source pair $Y_{A_{s}=63}/Y_{A_{s}=60}$ (squares),
$Y_{A_{s}=66}/Y_{A_{s}=63}$ (circles), $Y_{A_{s}=69}/Y_{A_{s}=66}$
(up-triangles), $Y_{A_{s}=66}/Y_{A_{s}=60}$ (down-triangles), and
$Y_{A_{s}=69}/Y_{A_{s}=63}$ (diamonds). }
\end{figure}

\begin{figure}
\includegraphics[width=0.5\textwidth]{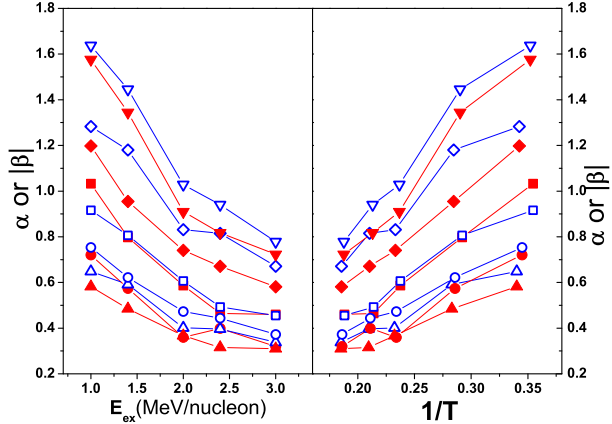}
\caption{\label{fig:fig5} (Color online) Same as Fig. 4 but
 for various source pairs with
$Z_s$ = 50. Symbols in the figure are  $\alpha$ (solid symbols) or
$|\beta|$ (open symbols)from source pair
$Y_{A_{s}=105}/Y_{A_{s}=100}$ (squares),
$Y_{A_{s}=110}/Y_{A_{s}=105}$ (circles),
$Y_{A_{s}=115}/Y_{A_{s}=110}$ (up-triangles),
$Y_{A_{s}=110}/Y_{A_{s}=100}$ (down-triangles), and
$Y_{A_{s}=115}/Y_{A_{s}=105}$ (diamonds).}
\end{figure}

\begin{table} \caption{\label{tab:table1} Level density parameters
$a_{T}$ and temperature $T$ for different source systems and
excitation energies.}
\begin{ruledtabular}
\begin{tabular}{lccccc}
Source&$E_{ex}$(MeV)&$a_{T}$&$T(MeV)\footnote{refers to the saddle-point configuration temperature}$& \\
\hline
$Z_s$=30 $A_s$=60/63& 1.0&7.5/7.8&2.8/2.9 \\
                    & 1.4&7.3/7.6&3.4/3.5 \\
                    & 2.0&7.1/7.4&4.1/4.2 \\
                    & 2.4&7.0/7.3&4.5/4.6 \\
                    & 3.0&6.8/7.1&5.1/5.2 \\

\hline
$Z_s$=30,$A_s$=66/69&1.0 &8.1/8.5&2.9/2.9 \\
                    &1.4 &8.0/8.3&3.5/3.5 \\
                    &2.0 &7.7/8.0&4.2/4.2 \\
                    &2.4 &7.6/7.8&4.6/4.7 \\
                    &3.0 &7.4/7.6&5.2/5.3 \\
\hline \hline
$Z_s$=50,$A_s$=100/105&1.0&11.8/12.3&2.8/2.9 \\
                      &1.4&11.5/11.9&3.4/3.5 \\
                      &2.0&11.0/11.4&4.2/4.3 \\
                      &2.4&10.8/11.2&4.6/4.7 \\
                      &3.0&10.5/10.9&5.3/5.4 \\
\hline
$Z_s$=50,$A_s$=110/115&1.0&12.8/13.2&2.9/3.0 \\
                      &1.4&12.4/12.8&3.5/3.6 \\
                      &2.0&11.9/12.3&4.3/4.3 \\
                      &2.4&11.6/12.0&4.8/4.8 \\
                      &3.0&11.3/11.8&5.4/5.4 \\
\end{tabular}
\end{ruledtabular}
\end{table}

The excitation energy and temperature dependences of $\alpha$ and
$\beta$ are shown in Fig. \ref{fig:fig4} and Fig. \ref{fig:fig5} for
different size and isospin asymmetry source pairs, respectively.
Temperature of the initial source are tabulated in TABLE
\ref{tab:table1}, in which the level density parameter $a_{T}$ is
derived from the Eq. (\ref{eq:eq6}) and the temperature are
calculated from Eq. (\ref{eq:eq9}).

Both parameters $\alpha$ and $|\beta|$ in Fig. \ref{fig:fig4} and
Fig. \ref{fig:fig5} have monotonic dependence on the excitation
energy and their absolute values decrease with the excitation
energy, $|\beta|$ generally are greater than $\alpha$. In right
panel of Fig. \ref{fig:fig4} and Fig. \ref{fig:fig5}, $\alpha$ and
$|\beta|$ show linear dependence on $1/T$, which evidences the
relationship of $\alpha = \Delta\mu_{n}/T$ and $\beta =
\Delta\mu_{p}/T$ \cite{Ts01b}. The slope of the relation between
$\alpha$ ($|\beta|$) and $1/T$ should be the free neutron (proton)
chemical potential difference $\Delta\mu_{n}$ ($|\Delta\mu_{p}|$),
the linear dependence of $\alpha$ ($|\beta|$) on $1/T$ also
evidences the constant of free neutron (proton) chemical potential
difference $\Delta\mu_{n}$ ($|\Delta\mu_{p}|$) between two initial
sources with defined asymmetry isospin, which are independent of the
excitation energy or temperature. The linear dependence of
isoscaling parameters $\alpha$ and $\beta$ with $1/T$ offers a
signal to calculate the free neutron (proton) chemical difference
$\Delta\mu_{n}$ ($|\Delta\mu_{p}|$), which can be calculated from
the slope fitting in the right panels of Fig. \ref{fig:fig4} and
Fig. \ref{fig:fig5}. Considering $\alpha>0$ and $\beta<0$, the free
neutron (proton) chemical potential difference $\Delta\mu_{n}>0$ and
$\Delta\mu_{p}<0$.

For an isospin asymmetry system $I=(N-Z)/A$, its neutron and
proton chemical potential are  $\mu_{n,I}$ and $\mu_{p,I}$,
respectively. From the isoscaling calculation we can get the free
neutron and proton chemical potential difference
$\Delta\mu_{n}=\mu_{n,I_2}-\mu_{n,I_1}$ and
$\Delta\mu_{p}=\mu_{p,I_2}-\mu_{p,I_1}$, thus
$\Delta\mu_{n}-\Delta\mu_{n}=(\mu_{n,I_2}-\mu_{n,I_1})-(\mu_{p,I_2}-\mu_{p,I_1})$.
If one isospin symmetry system with $I=0$ is included,
$\mu_{n,I=0}=\mu_{p,I=0}$, the difference of free neutron and
proton chemical potential $\mu_{nI}-\mu_{pI}$ of the isospin
asymmetry system can be deduced as shown in \cite{Bo02, Ba05}.
This $\mu_{nI}-\mu_{pI}$ extracted from Fig. \ref{fig:fig4} and
\ref{fig:fig5} are printed in TABLE \ref{tab:table2}, in which the
index of isospin asymmetry parameter $I$ was replaced by the mass
number $A_{s}$.

\begin{table}
\caption{\label{tab:table2} Difference of free neutron and proton
chemical potential $\mu_{nI}-\mu_{pI}$ for series isospin asymmetry
systems from the slope fitting in Fig. \ref{fig:fig4} and
\ref{fig:fig5}.}
\begin{ruledtabular}
\begin{tabular}{cccc}
System& $\Delta\mu_{n}$\footnote{refers to
$\mu_{n,A_{s_2}}-\mu_{n,A_{s_1}}$}(MeV) &
$\Delta\mu_{p}$\footnote{refers to
$\mu_{p,A_{s_2}}-\mu_{p,A_{s_1}}$}(MeV)&
$\mu_{n}-\mu_{p}$\footnote{refers to
$(\mu_{n,A_{s_2}}-\mu_{p,A_{s_2}})-(\mu_{n,A_{s_1}}-\mu_{p,A_{s_1}})$}(MeV) \\
$Z_s$ $A_{s_1}$ $A_{s_2}$ & & & \\
\hline
30  60   63 &2.79$\pm$0.21&3.42$\pm$0.33&6.21$\pm$0.39 \\
30  63   66 &2.94$\pm$0.29&3.53$\pm$0.30&6.47$\pm$0.42 \\
30  66   69 &1.64$\pm$0.24&2.42$\pm$0.53&4.06$\pm$0.58 \\
30  60   66 &5.38$\pm$0.23&5.77$\pm$0.49&11.25$\pm$0.54 \\
30  63   69 &4.35$\pm$0.20&4.71$\pm$0.22&9.06$\pm$0.30 \\
\hline
50  100  105&3.66$\pm0.28$&2.96$\pm$0.29&6.62$\pm$0.40 \\
50  105  110&2.58$\pm$0.35&2.40$\pm$0.10&4.98$\pm$0.36 \\
50  110  115&1.89$\pm$0.15&2.13$\pm$0.29&4.02$\pm$0.33 \\
50  100  110&5.53$\pm$0.48&5.41$\pm$0.49&10.94$\pm$0.69 \\
50  105  115&3.95$\pm$0.11&4.06$\pm$0.53&8.01$\pm$0.54 \\
\end{tabular}
\end{ruledtabular}
\end{table}

\subsection{Dependences of $\alpha$ and $\beta$ on isospin asymmetry and source size }

\begin{figure}
\includegraphics[width=0.5\textwidth]{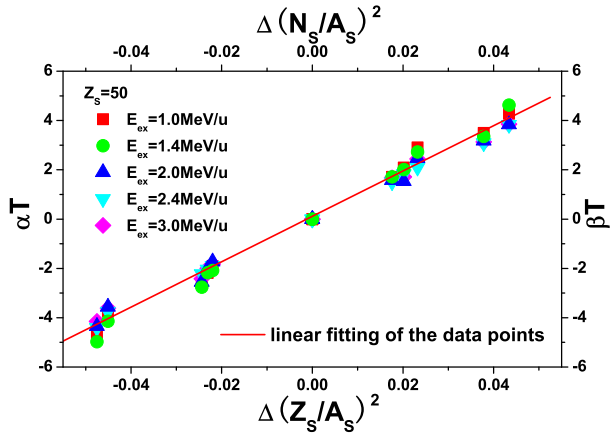}
\caption{\label{fig:fig6} (Color online) $\alpha\cdot T$ (positive
parts) and $\beta\cdot T$ (negative parts) as a function of
$(Z_s/A_s)_{1}^2-(Z_s/A_s)_{2}^2$ or
$(N_s/A_s)_{1}^2-(N_s/A_s)_{2}^2$ of the sources for various source
pairs with atomic number $Z_s = 50$ which is illustrated in the
figure at different excitation energies. }
\end{figure}

\begin{figure}
\includegraphics[width=0.5\textwidth]{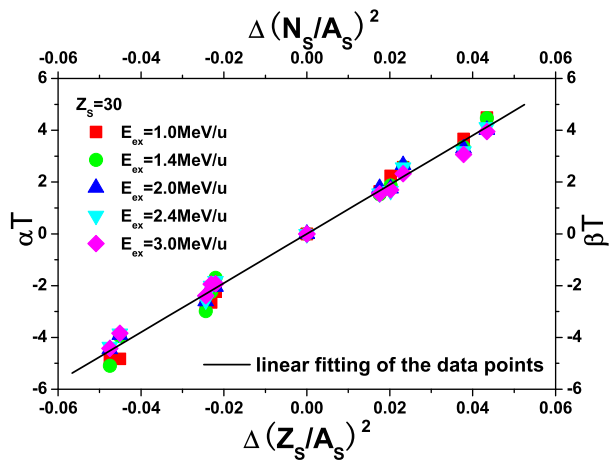}
\caption{\label{fig:fig7} (Color online) Same as Fig. 6 but for
various source pairs with  $Z_s = 30$ which is illustrated in the
figure at different excitation energies. }
\end{figure}

To explore the origin of isoscaling behavior and the dependence of
isoscaling parameters $\alpha$ and $\beta$ on the system size and
the isospin composition, we performed calculations on source
systems with different sizes and asymmetry ($N_s/Z_s$) values.
Symmetry energy coefficient $C_{sym}$ is dependent on not only the
nuclear density $\rho$, but also the system temperature
\cite{Li06}. In GEMINI investigation, the nuclear density is set
to be around the saturation density $\rho_{0}$ so that the system
density influences on the extraction of symmetry energy
coefficient can be neglected in this case. The symmetry energy
coefficient $C_{sym}$ can be seen only temperature dependent.
Since the systems have different sizes, the temperatures of the
system are slightly different even though the excitation energy is
fixed at same values. In this case, in Fig. \ref{fig:fig6} and
\ref{fig:fig7} the temperature is served as a correction factor,
we made the temperature correction by $\alpha\cdot T$ and
$\beta\cdot T$. Fig. \ref{fig:fig6} and \ref{fig:fig7} depicts
$\alpha\cdot T$ as a function of
$(Z_{s}/A_{s})_{1}^{2}-(Z_{s}/A_{s})_{2}^{2}$  and $\beta\cdot T$
as a function of $(N_{s}/A_{s})_{1}^{2}-(N_{s}/A_{s})_{s}^{2})$ of
the initial source pair, plotted in the same figure for different
excitation energies. The source pairs we simulated are $Z_s$ = 50
pairs ($Z_s$ = 50: $A_s$ = 100, 105, 110 and 115) and $Z_s$ = 30
pairs ($Z_s$ = 30: $A_s$ = 60, 63, 66 and 69). All these systems
with different source sizes and isospin asymmetries lie along one
line, which illustrates that the isoscaling parameters $\alpha$
and $\beta$ are not sensitive to the system size and charge. The
linear fits of the calculation points in Fig. \ref{fig:fig6} and
\ref{fig:fig7} are printed on the figure, slopes in Fig.
\ref{fig:fig6} are $92.0\pm2.8$ and $95.0\pm1.6$ in Fig.
\ref{fig:fig7}, respectively. This approximate linear relationship
has been also observed in many experimental data \cite{Bo02, So03}
as well as other model calculations with EES, SMM model
\cite{Ts01b}, AMD \cite{Sh04} and IQMD model \cite{Ti05}.

It has been shown in the framework of the grand-canonical limit of
the statistical multifragmentation model \cite{Ts01b} and in the
expanding-emitting source model \cite{Bo02}, a simple relationship
can be derived between isoscaling parameters $\alpha$ and
$\Delta(Z_s/A_s)^2$, $\beta$ and$\Delta(N_s/A_s)^2$ of the sources
by following forms:

\begin{equation}
\alpha =
4\frac{C_{sym}}{T}\Big[\Big(\frac{Z_{s}}{A_{s}}\Big)_1^2-\Big(\frac{Z_{s}}{A_{s}}\Big)_2^2\Big]
\label{eq:eq10}
\end{equation}
and
\begin{equation} \beta =
4\frac{C_{sym}}{T}\Big[\Big(\frac{N_{s}}{A_{s}}\Big)_1^2-\Big(\frac{N_{s}}{A_{s}}\Big)_2^2\Big]
. \label{eq:eq11}
\end{equation}

Above equations have been proved to be good approximations by many
calculations and experimental data, and are generally adopted to
constrain the symmetry energy coefficient $C_{sym}$ in experiment.
Here they are verified by the results in Fig. \ref{fig:fig6} and
\ref{fig:fig7}, i.e. the slope of $\alpha\cdot T$ with respect to
$\Delta(Z_{s}/A_{s})^2$ and $\beta\cdot T$ with respect to
$\Delta(N_{s}/A_{s})^2$. By linear fitting the data in Fig.
\ref{fig:fig6} and \ref{fig:fig7} with Eq. (\ref{eq:eq10}) and
(\ref{eq:eq11}) the symmetry energy coefficient $C_{sym}$ can be
constrained to be $23.0\pm0.7 MeV$ from Fig. \ref{fig:fig6} and
$23.8\pm0.4 MeV$ from Fig. \ref{fig:fig7}, comparable with the
values from other models and experiments \cite{Bo02, So03, Ts01b,
Sh04}.

\section{\label{sec:sec4}Surface contribution to isospin effect}

GEMINI code involves the isospin effect or symmetry energy in
calculating the binding energy. For heavy systems ($Z_s > $12),
the mass excess of the initial and residual systems of a spherical
nucleus is given by following equation \cite{Kr79, Mo81}

\begin{figure}
\includegraphics[width=0.5\textwidth]{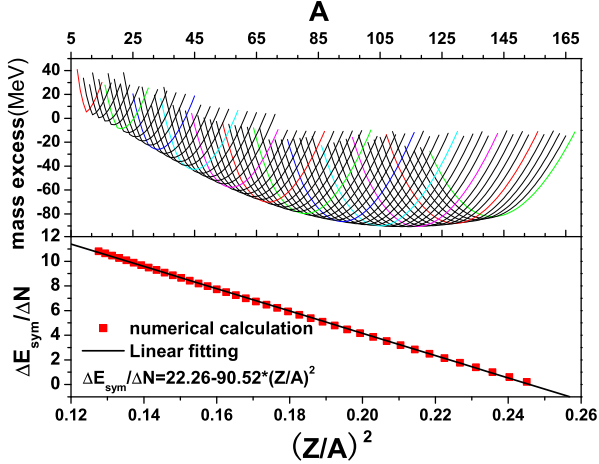}
\caption{\label{fig:fig8} (Color online) Top panel: mass excess
calculated from Eq. (\ref{eq:eq12}) as a function of nuclear mass
number, curves from left to right correspond to the atomic number
$Z$ of isotopes series from $Z$ = 5 to $Z$ = 60 in order; Bottom
panel: numerical calculated derivative of symmetry energy $E_{sym}$
(MeV/nucleon) with respect to neutron number $N$ in Eq.
(\ref{eq:eq12}) as a function of $(Z/A)^2$ (solid marks) and the
linear fit of the numerical calculation, fitted result is printed on
the Figure.}
\end{figure}

\begin{widetext}
\begin{eqnarray}
M_{macro}^{(0)}&=&M_{n}N+M_{p}Z-a_{v}(1-k_{v}I^{2})A+a_{s}(1-k_{s}I^{2}) \nonumber \\
&\times&\left\{A^{2/3}-3\left(\frac{a}{r_{0}}\right)^2
+\left(\frac{r_{0}}{a}A^{1/3}+1\right)
\left[2A^{2/3}+3\frac{a}{r_{0}}A^{1/3}+3\left(\frac{a}{r_{0}}\right)^2\right]
e^{2r_{0}A^{1/3}/a}\right\} \nonumber \\
&+&\frac{3}{5}\frac{e^2}{r_{0}}\left[\frac{Z^2}{A^{1/3}}
-\frac{5}{2}\left(\frac{b}{r_{0}}\right)^{2}\frac{Z^2}{A}
-\frac{5}{4}\left(\frac{3}{2\pi}\right)^{2/3}\frac{Z^{4/3}}{A^{1/3}}\right]
+W(|I|+d)-a_{el}Z^{2.39}+
\begin{cases}
\Delta-\frac{1}{2}\delta,& \text{N and Z odd} \\
\frac{1}{2}\delta,& \text{N or Z odd} \\
-(\Delta-\frac{1}{2}\delta),& \text{N and Z even}
\end{cases}
\nonumber \\
\label{eq:eq12}
\end{eqnarray}
\end{widetext}
where $I = (N-Z)/A$. The parameters are taken from \cite{Mo81},
$M_{n}$ = 8.071431, $M_{p}$ = 7.289034, $e^{2}$ = 1.4399764, $b$ =
0.99, $W$ = 36, $a_{el}$ = 1.433$e^{-5}$, $r_{0}$ = 1.16, $a$ =
0.68, $a_{s}$ = 21.13, $k_{s}$ = 2.3, $a_{v}$ = 15.9937, $k_{v}$ =
1.927, $\Delta$ = 12/$\sqrt A$, $\delta$ = 20/A. Correction
arising from single-particle effect is added to Eq.
(\ref{eq:eq12}), which includes Coulomb energy for diffusive
surface, proton form factor and charge asymmetry term \cite{Kr79,
Mo81, Da76}. In the top panel of Fig. \ref{fig:fig8}, the
calculated mass excess by Eq. (\ref{eq:eq12}) is plotted as a
function of mass number $A$, the isotopes with different charge
number $Z$ from 5 to 60 are presented by lines from left to right.

In Eq. (\ref{eq:eq12}) the dominant symmetry energy term arises from
two parts: the volume symmetry energy part $a_{v}k_{v}I^{2}A$ and
the surface diffuseness term related with the isospin asymmetry
$-a_{s}k_{s}I^{2}\left\{A^{2/3}+......\right\}$. As derived in other
models and theoretical frame \cite{Ts01, Ts01b}, isotope yield ratio
is dominantly determined by the symmetry term in the binding energy
for two equilibrium sources with comparable mass and temperature but
different isospin degree $N_{s}/Z_{s}$. In this case, the isoscaling
parameter $\alpha$ can be achieved by following approximate form
\begin{equation}
\alpha=-\Delta s_{n}/T,  \label{eq:eq13}
\end{equation}
where $\Delta s_{n}$ is the difference in neutron separation
energy between the two sources, considering the dominant term in
separation energy is symmetry term, which is calculated in GEMINI
simulation by Eq. (\ref{eq:eq13}). The symmetry term taken from
Eq. (\ref{eq:eq13}) can be expressed alone by

\begin{equation}
E_{sym}=c_{v}I^{2}A-c_{s}I^{2}\left\{A^{2/3}+......\right\},
\label{eq:eq14}
\end{equation}
the first term in Eq. (\ref{eq:eq14}) is the volume term of
isospin asymmetry part, the second term is the surface effect of
isospin asymmetry. $c_{v} = a_{v}k_{v} = 15.9937 \times  1.927 =
30.8199$ MeV which is the generally used symmetry energy
coefficient at saturate nuclear density $\rho_{0}$=0.16 $fm^{-3}$,
$c_{s} = a_{s}k_{s} = 21.13 \times 2.3 = 48.599$ MeV. The
difference in neutron separation energy between two sources can be
approximately obtained by taking the derivatives of the symmetry
energy of Eq. (\ref{eq:eq14}) with respect to $N$. Numerical
calculation of the derivatives of Eq. (\ref{eq:eq14}) with respect
to $N$ (i.e. $\Delta E_{sym}/\Delta N$) is performed, and
approximate linear function on $(Z/A)^2$ is observed, shown in the
bottom panel of Fig. \ref{fig:fig8}. The linear fits give $\Delta
E_{sym}/\Delta N = 22.26019-90.5205(Z/A)^{2} \approx
22.63(1-4(Z/A)^2)$, effective symmetry energy coefficient
$C_{sym}$ in the equation of state deduced here approximates 22.63
MeV, which is consistent with the symmetry energy coefficient
$C_{sym}$ derived from isoscaling parameters $\alpha$ and $\beta$
very well, namely $C_{sym} = 23.0\pm0.7$ MeV derived from $\alpha$
or $C_{sym} = 23.8\pm0.4$ MeV from $\beta$. As we know, the
derivatives with respect to $N$ of the volume term in Eq.
(\ref{eq:eq14}) is $c_{v}(1-4(Z/A)^2) = 30.8199-123.2796(Z/A)^{2}
= 30.8199(1-4(Z/A)^2)$, hence the contribution from the surface
effect is approximately $22.630125(1-4(Z/A)^2)
-30.8199(1-4(Z/A)^2) $ = $-8.55971(1-4(Z/A)^2)$. Sum of the volume
and surface terms of the symmetry energy gives the consistent
result with the experimental results \cite{Ts01b} and other model
calculations, which indicates that the surface effect term of the
symmetry energy strongly influences the isoscaling parameter
$\alpha$ and $\beta$.

\section{\label{sec:sec5}Summary}

The statistical sequential decay process of the equilibrated
source has been successfully investigated by the GEMINI model. The
model constrains the source size and fragments density at saturate
nuclear density $\rho_0$ and the decay starts from an equilibrated
system which is separated from any dynamical process. By only
calculating the first step decay, the source size, density and
temperature can be well defined. In our work we apply it to survey
the evolution of isospin degree of freedom. Isospin effects on the
decayed fragments from different source sizes, isospin asymmetries
and excitation energies have been systematically investigated.
Isoscaling phenomena are observed for the emission fragments of
light fragments. Isoscaling parameters $\alpha$ and $|\beta|$
decrease with the increasing of excitation energy, they generally
show linear dependence on the inverse of the system temperature
$T$, and are independent of the source size. From the linear
function between $\alpha$ versus 1/T or $\beta$ versus 1/T, the
extracted linear slope is the free neutron (proton) chemical
potential difference $\Delta\mu_{n}$ and $\Delta\mu_{p}$ between
two systems. If the isospin symmetry source is included, then the
difference of neutron and proton chemical potential
$\Delta\mu_{n}-\Delta\mu_{p}$ can be determined, thus the slope
difference between $\alpha$ and $\beta$ versus $1/T$ offer a
signal to calculate the neutron and proton chemical potential
difference. After considering system temperature, the simple
relationship $\alpha =
4C_{sym}[(Z_{1}/A_{1})^{2}-(Z_{2}/A_{2})^{2}]/T$ and $\beta =
4C_{sym}[(N_{1}/A_{1})^{2}-(N_{2}/A_{2})^{2}]/T$ can be well
reproduced for various sources. Only when the surface effect term
in the symmetry energy is taken into accounted, the above linear
relationship between $\alpha$ and $\Delta(Z/A)_s^2$, or $\beta$
and $\Delta(N/A)_s^2$ can give consistent  symmetry energy
coefficient $C_{sym}$ with experimentally proposed results and
other model results, which illustrates that the surface effect
plays a significant role in the symmetry energy term.

\vspace{3mm}

This work was supported in part by the National Natural Science
Foundation of China under Grant No 10405033, 10505026, 10405032,
10535010, 10610285, the Shanghai Development Foundation for
Science and Technology under Grant Numbers 05XD14021 and
06JC14082, the Knowledge Innovation Project of Chinese Academy of
Sciences under Grant No.  KJXC3-SYW-N2, Agreement of Scientific
Cooperation between China and Slovakia by Ministry of Sciences and
Technology.


\end{document}